\documentclass[notitlepage,superscriptaddress,showpacs,nobalancelastpage,twocolumn,aps,prl,floatfix,longbibliography]{revtex4-2}
\usepackage[english]{babel}
\RequirePackage[T1]{fontenc}
\RequirePackage{times}
\usepackage{siunitx}
\usepackage[italicdiff]{physics}
\usepackage{amsfonts}
\usepackage{amsmath}
\usepackage{amssymb}
\usepackage{graphicx,epstopdf}
\usepackage{appendix}
\usepackage{xspace}
\usepackage{array}
\usepackage[hidelinks]{hyperref}
\usepackage{color}
\usepackage{xcolor}
\usepackage{bm}
\usepackage{verbatim}
\usepackage{tikz}
\usepackage{float}
\allowdisplaybreaks[3]

\begin{document}

\title{Prediction of antimagnon-polarons with intrinsic momentum-space level attraction}

\author{Chongzhou Wang}
\affiliation{State Key Laboratory of Surface Physics and Institute for Nanoelectronic Devices and Quantum Computing, Fudan University, Shanghai 200433, China}
\affiliation{Department of Physics, Fudan University, Shanghai 200433, China}

\author{Zhengda Jing}
\affiliation{State Key Laboratory of Surface Physics and Institute for Nanoelectronic Devices and Quantum Computing, Fudan University, Shanghai 200433, China}
\affiliation{Department of Physics, Fudan University, Shanghai 200433, China}

\author{Weichao Yu}
\email{wcyu@fudan.edu.cn}
\affiliation{State Key Laboratory of Surface Physics and Institute for Nanoelectronic Devices and Quantum Computing, Fudan University, Shanghai 200433, China}
\affiliation{Zhangjiang Fudan International Innovation Center, Fudan University, Shanghai 201210, China}

\date{\today}

\begin{abstract}
Propagating hybrid quasiparticles formed by magnons and phonons (magnon-polarons) or photons (magnon-polaritons) typically exhibit level repulsion in momentum space. Level attraction, by contrast, requires an anti-phase channel and has so far been restricted to zero-dimensional systems mediated by dissipative baths. Here we predict the antimagnon-polaron in a continuous magnetoelastic medium, which inherently exhibits momentum-space level attraction. The spin-orbit-torque (SOT)-stabilized antimagnon precesses with opposite handedness, directly reversing the magnetoelastic coupling without requiring an external environment. This intrinsic attraction yields a distinctive \(0/\pi\) phase-locked simultaneous growth. Micromagnetic simulations confirm this picture, revealing that the SOT-controlled balance between gain and damping drives a collective instability. This ultimately enables a continuously tunable, dc-pumped, single-frequency phonon laser, establishing a novel route for coherent acoustic emission.
\end{abstract}

\maketitle

\textit{Introduction}---Collective spin excitations (magnons) hybridize with lattice vibrations (phonons) to form propagating magnon-polarons, mixed quasiparticles central to spintronics and magnomechanics~\cite{kittel_interaction_1958,bozhko_magnon-phonon_2020}. Similar coherent hybridization is ubiquitous in other systems, including magnon-polaritons formed by magnons and photons, as well as exciton-polaritons. In these platforms, coupling between two coherent modes typically produces an avoided crossing of the hybridized branches. Close to resonance, the branches anticross, and the hybridization is said to display \textit{level repulsion}~\cite{man_direct_2017,flebus_magnon-polaron_2017}. This anticrossing redistributes spectral weight, polarization, and group velocity between the branches, thereby governing response and transport in diverse phenomena such as spin- and heat-transport anomalies~\cite{uchida_long-range_2011,weiler_spin_2012,kikkawa_magnon_2016}, acoustic driving and read-out of magnetization~\cite{lyons_acoustically_2023,ourdani_giant_2026,hwang_strongly_2024,wang_effective_2026}, band-structure engineering~\cite{kunstle_magnon-polaron_2025,liao_hybrid_2024,hatanaka_phononic_2023}, and nonreciprocal spin-wave transport~\cite{yamamoto_non-reciprocal_2020,tateno_highly_2020}. Fundamentally, this ubiquitous level repulsion originates from the in-phase nature of the coupling, which mathematically yields a positive coupling term that forces the eigenfrequencies to repel.

\textit{Level attraction} was first discovered in cavity magnonic systems~\cite{harder_level_2018}. In most subsequent realizations, this attraction is fundamentally driven by dissipative coupling, where the coupling between two modes is mediated by a shared dissipative bath that establishes an anti-phase interaction channel. Under such dissipative coupling, the branches cease to avoid each other and bend \emph{towards} one another, coalescing in frequency and linewidth~\cite{rao_level_2019,wang_dissipative_2020,tserkovnyak_exceptional_2020}. Attraction provides properties that repulsion cannot, including linewidth narrowing, enhanced sensitivity near the coalescence, and the ability to generate lasing from two passive modes~\cite{zhang_dissipative_2022,wang_single-mode_2025}. However, these reported phonon lasers rely on an external dissipative environment to mediate the coupling. Experimentally, these observations have so far been restricted to zero-dimensional systems, where the relevant magnetic mode is a localized, non-propagating magnon with \(k=0\), typically the Kittel mode or a higher-order localized magnon mode~\cite{yao_microscopic_2019,yang_control_2019,gardin_level_2025,bernier_level_2018,lu_level_2023,todaka_phase_2026}. To date, momentum-space level attraction has not been reported for propagating hybrids such as magnon-polarons or magnon-polaritons in a continuous medium.

Here we predict the antimagnon-polaron, a propagating hybrid exhibiting momentum-space level attraction in a continuous magnetoelastic medium. Unlike the environment-mediated dissipative coupling that requires a shared bath~\cite{zhang_dissipative_2022}, the antimagnon provides an intrinsic anti-phase channel to realize this attraction. Stabilized by a dc spin--orbit torque (SOT)~\cite{harms_antimagnonics_2024}, the antimagnon precesses with opposite handedness about an inverted equilibrium. This handedness flip causes the background magnetization to enter the magnetoelastic coupling vertex with an opposite sign, directly reversing the coupling term in the secular equation. Bassant \emph{et al.} first investigated the hybridization of antimagnons and phonons in bulk and applied it to interfacial phonon amplification~\cite{bassant_phonon_2025}. Here we analyze the complex spectrum of the continuous medium and identify its intrinsic momentum-space level attraction. Very recently, Wang \emph{et al.} provided direct Brillouin-light-scattering evidence of antimagnons with inverted dispersion in a compensated Pt/BiYIG film~\cite{wang_direct_2026}, establishing the propagating branch. Building on this progress, we explore the coherent hybridization of this branch with phonons. We further show that the SOT governs the stability of the resulting hybrid mode, identify the in-phase time-domain fingerprint of the attraction channel, and propose a dc-pumped phonon laser based on this gain mechanism.

\begin{figure}[t]
    \centering
    \includegraphics[width=\columnwidth]{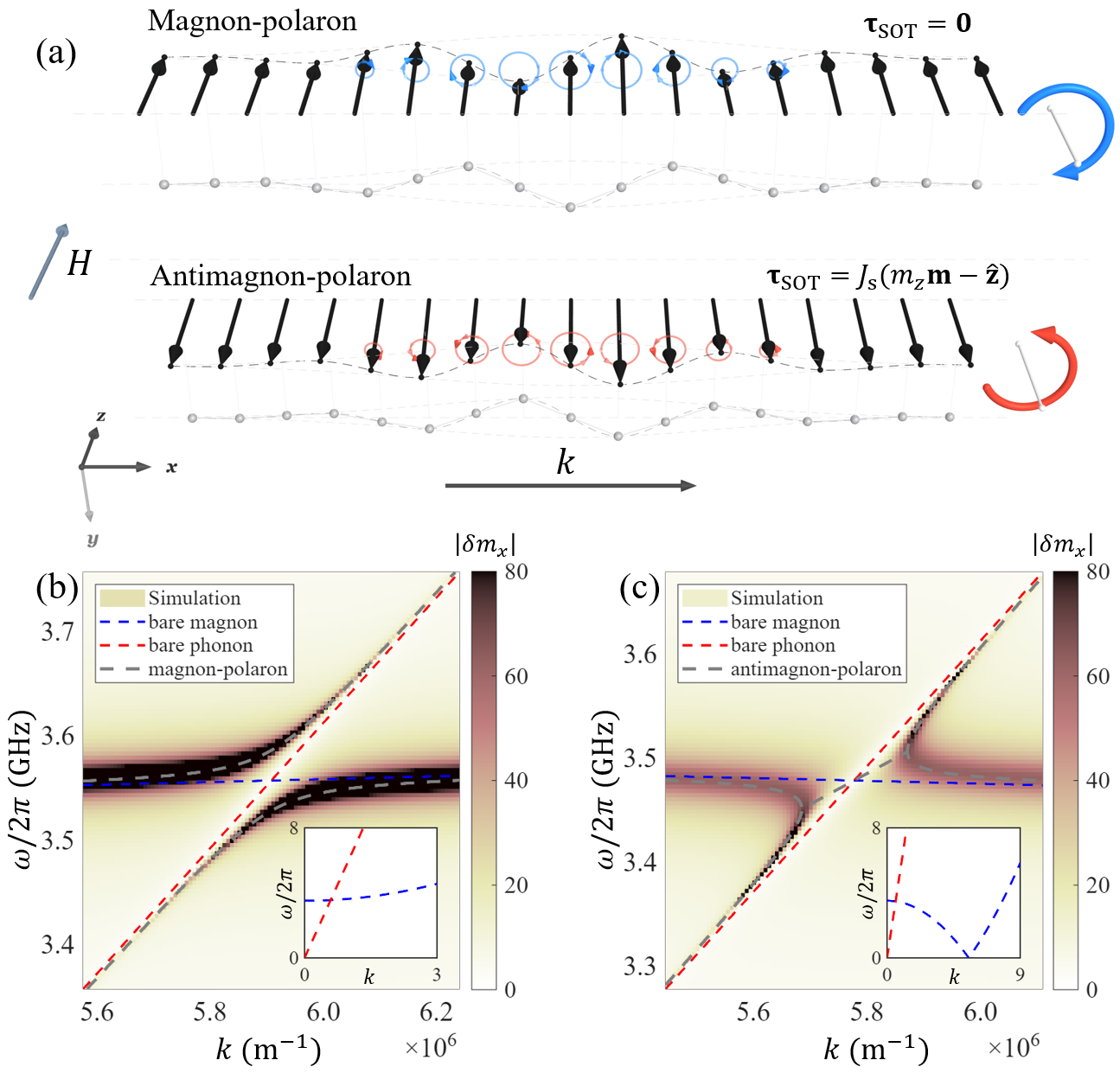}
    \caption{Concept and simulated dispersions of the antimagnon-polaron. (a) Schematics of the magnon-polaron (top, SOT off) with right-handed precession, and the antimagnon-polaron (bottom, SOT on) where the SOT stabilizes the antiparallel equilibrium, resulting in left-handed precession. (b),(c) Simulated spectral function $|\delta m_x|$ versus frequency $\omega/2\pi$ and wavevector $k$. The magnon-polaron (b) exhibits level repulsion (avoided crossing), while the antimagnon-polaron (c) exhibits level attraction, coalescing near the crossing point. Blue and red dashed lines show the bare magnon and phonon dispersions, and gray dashed lines show the hybridized branches. Insets: zoom-out views showing the full range of the bare phonon and (anti)magnon modes.}
    \label{fig:fig1}
\end{figure}

\textit{Model and microscopic origin}---We consider a one-dimensional magnetoelastic system where the unit magnetization $\mathbf m=\mathbf M/M_s$ obeys the Landau-Lifshitz-Gilbert equation augmented by an SOT with polarization $\mathbf p=-\hat{\mathbf{z}}$~\cite{fert_electrical_2024,harms_dynamically_2021,zhou_left-handed_2019},
\begin{equation}
\frac{\partial\mathbf m}{\partial t}
=-\gamma\,\mathbf m\times\mathbf H_{\rm eff}
+\alpha\,\mathbf m\times\frac{\partial\mathbf m}{\partial t}
+J_{\mathrm{s}}\,\mathbf m\times(\mathbf p\times\mathbf m),
\label{eq:LLG}
\end{equation}
where $\mathbf M$ is the magnetization, $M_s$ the saturation magnetization, $\gamma$ the gyromagnetic ratio, $\alpha$ the Gilbert damping, and $J_{\mathrm{s}}>0$ the SOT rate. The effective field contains exchange, the static external field $\mathbf H_0=H\hat{\mathbf{z}}$, and the magnetoelastic contribution, $\mathbf H_{\rm eff}=A\nabla^2\mathbf m+\mathbf H_0+\mathbf H^{\rm mec}$, where \(A\) is the exchange coefficient, with the cubic magnetoelastic energy density $U_{\rm mec}=b_1\sum_i m_i^2\varepsilon_{ii}+b_2\sum_{i\neq j}m_im_j\varepsilon_{ij}$~\cite{kittel_interaction_1958}, where \(b_1\) and \(b_2\) are the magnetoelastic constants, and $\varepsilon_{ij}=\tfrac12(\partial_iu_j+\partial_ju_i)$. The displacement $\mathbf u$ obeys the elastic wave equation $\rho\ddot{\mathbf u}+\beta\dot{\mathbf u}=\mu\nabla^2\mathbf u+(\lambda+\mu)\nabla(\nabla\cdot\mathbf u)+\mathbf f^{\rm mec}$, where \(\beta\) is the phonon damping coefficient, with the magnetoelastic body force $\mathbf f^{\rm mec}=\nabla\cdot\bar\Sigma$, $\Sigma_{ij}=\partial U_{\rm mec}/\partial(\partial_ju_i)$. The full model is summarized in the Supplemental Material (SM)~\cite{sm}.

\begin{figure*}[t]
    \centering
    \includegraphics[width=\textwidth]{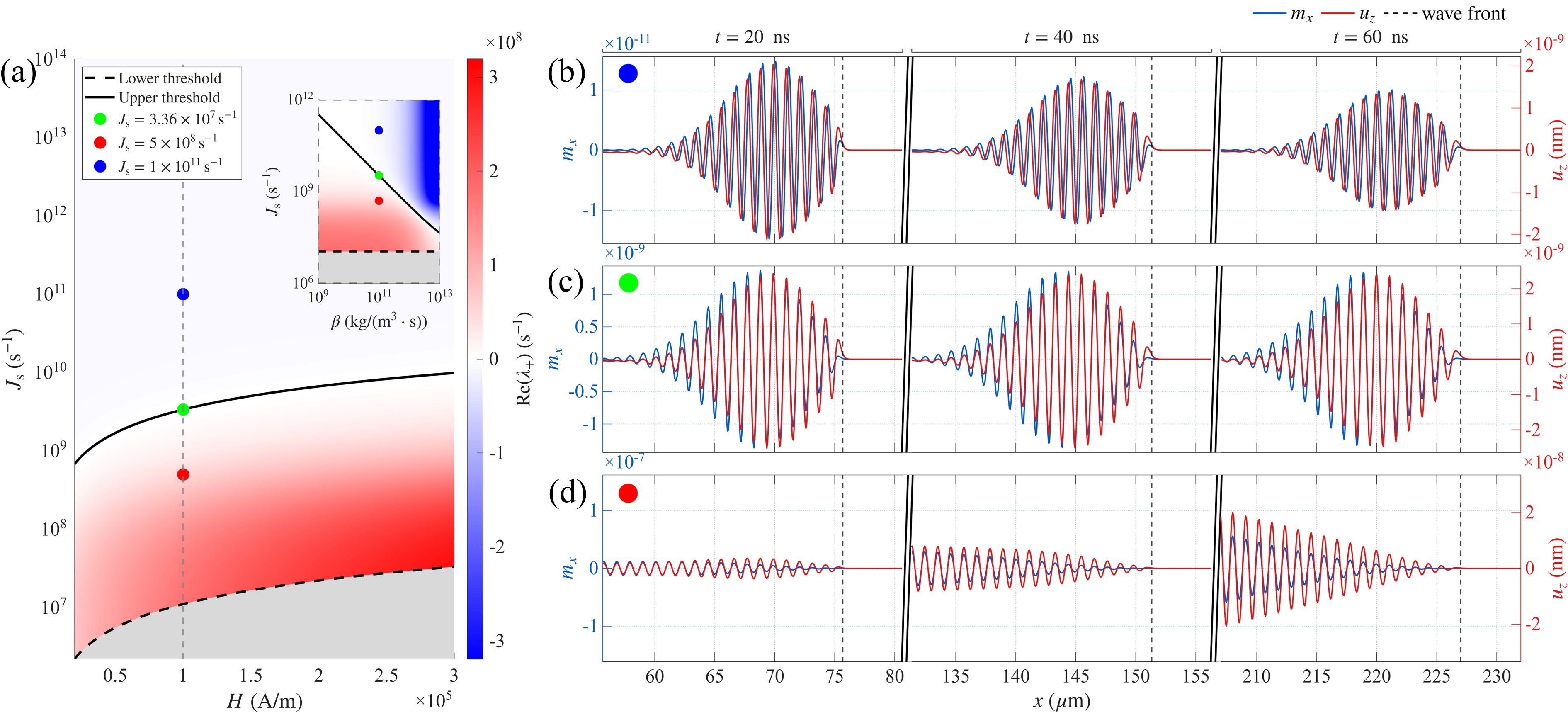}
    \caption{Stability phase diagram and time-domain fingerprints of the antimagnon-polaron. (a) Dominant growth rate $\mathrm{Re}\,\lambda_+$ in the $(H, J_{\mathrm{s}})$ plane, calculated from Eq.~\eqref{eq:growth}. The dashed and solid black lines mark the lower ($J_{\mathrm{s}}^{\rm st}=\alpha\gamma H$) and upper ($J_{\mathrm{s}}^{\rm crit}$) stability thresholds, respectively. The white region denotes the growth window, while the grey region marks the regime below the stability threshold, where the inverted equilibrium cannot form and the antimagnon cannot be excited, resulting in the decay of both hybrid modes. Green, red, and blue dots mark the SOT rates $J_{\mathrm{s}}=3.36\times10^{9}$, $5\times10^{8}$, and $1\times10^{11}$~s$^{-1}$ used in (b)--(d). Inset: phase diagram in the $(J_{\mathrm{s}},\beta)$ plane at fixed $H$, showing the same thresholds. (b)--(d) Micromagnetic simulations of the spatiotemporal evolution of $m_x$ (blue, left axis) and $u_z$ (red, right axis) at $t=20$, $40$, and $60$~ns for the three SOT rates marked in (a). Dashed vertical lines in (b)--(d) mark the ballistic acoustic wave fronts. At $J_{\mathrm{s}}=3.36\times10^{9}$~s$^{-1}$ (b), the magnetization and displacement oscillate in phase with large amplitude. At $J_{\mathrm{s}}=5\times10^{8}$~s$^{-1}$ (c), the in-phase relation is preserved but the amplitude is reduced. At $J_{\mathrm{s}}=1\times10^{11}$~s$^{-1}$ (d), the coupled antimagnon-polaron mode is strongly suppressed, consistent with the SOT double-edged-sword picture.}
    \label{fig:fig2}
\end{figure*}

Throughout the numerical simulations presented below, we adopt the standard yttrium iron garnet (YIG) material parameters~\cite{clark_elastic_1961,comstock_magnetoelastic_1965}: mass density $\rho=5170$~kg/m$^3$, shear modulus $\mu=74$~GPa, Lam\'e constant $\lambda=104$~GPa, transverse sound velocity $c_t\approx3783$~m/s, gyromagnetic ratio $\gamma=2.21\times10^5$~rad\,s$^{-1}$(A/m)$^{-1}$, saturation magnetization $M_s=0.194\times10^6$~A/m, Gilbert damping $\alpha=5\times10^{-4}$, exchange coefficient $A=0.328\times10^{-10}$~A$\cdot$m, magnetoelastic constants $b_1=3.48\times10^5$~J/m$^3$, $b_2=2b_1$, and effective magnetoelastic coupling $b\equiv b_2/(\mu_0M_s)\approx2.86\times10^6$~A/m. The external static field is fixed at $H=1.0\times10^5$~A/m. All micromagnetic simulations are performed using the Micromagnetics Module developed for COMSOL Multiphysics~\cite{comsol,Zhang2023,wang_effective_2026}.

The SOT stabilizes an antiparallel equilibrium $\mathbf m_0=-\hat{\mathbf{z}}$ ($\sigma=-1$), in contrast to the conventional parallel state $\mathbf m_0=+\hat{\mathbf{z}}$ ($\sigma=+1$)~\cite{bassant_phonon_2025,harms_dynamically_2021}. We now linearize the coupled system around $\mathbf m_0=\sigma\hat{\mathbf{z}}$ and work with the phasor convention $\delta\mathbf m,\delta\mathbf u\propto e^{i\omega t}$, so that $\operatorname{Im}\omega>0$ corresponds to decay. For a plane wave of wavevector $k$, the spin sector reads
\begin{align}
i\omega\,\delta m_x&=-\Omega_\sigma\delta m_y-i\omega\alpha\sigma\,\delta m_y+\sigma J_{\mathrm{s}}\,\delta m_x,\notag\\
i\omega\,\delta m_y&=+\Omega_\sigma\delta m_x+i\omega\alpha\sigma\,\delta m_x+\sigma J_{\mathrm{s}}\,\delta m_y-i\gamma bk\,u_z,
\label{eq:linspin}
\end{align}
where $\Omega_\sigma(k)\equiv\omega_H+\sigma\omega_{\rm ex}(k)$, with $\omega_H=\gamma H$ and $\omega_{\rm ex}(k)=\gamma Ak^2$, and $b\equiv b_2/(\mu_0M_s)$. Neglecting acoustic damping, the elastic sector, with $\partial_x\to -ik$, becomes
\begin{equation}
(\omega_P^2-\omega^2)u_z=-i\sigma\frac{b_2}{\rho}k\,\delta m_x.
\label{eq:linelastic}
\end{equation}
with \(\omega_P=c_tk\) and $c_t=\sqrt{\mu/\rho}$. Eliminating $\delta m_y$ and applying the rotating-wave approximation (RWA) near the crossing $\omega\approx\omega_k^{(\sigma)}\approx\omega_P$, we obtain
\begin{align}
(\omega-\omega_k^{(\sigma)})\delta m_x&=-\frac{i\gamma bk}{2}u_z,
\label{eq:RWA1}\\
(\omega-\omega_P)u_z&=\frac{i\sigma b_2k}{2\rho\omega_P}\delta m_x,
\label{eq:RWA2}
\end{align}
with the bare spin-wave frequencies
\begin{equation}
\omega_k^{(\sigma)}\simeq\gamma(H+\sigma Ak^2)+i\sigma\big[\alpha\gamma(H+\sigma Ak^2)-J_{\mathrm{s}}\big].
\label{eq:bare}
\end{equation}
For $\sigma=+1$, Eq.~\eqref{eq:bare} is the ordinary magnon (denoted as \(\omega_k^{(m)}\)). For $\sigma=-1$, it is the antimagnon (denoted as \(\omega_k^{(a)}\)). The uncoupled antiparallel state is globally stable for $J_{\mathrm{s}}>\alpha\gamma H$ when $\omega_{\rm ex}(k)\ge0$, while the positive-frequency antimagnon branch requires \(|k|<\sqrt{H/A}\). At the crossing wavevector $k^*$, the weaker local condition is $J_{\mathrm{s}}>\alpha\gamma(H-Ak^{*2})$. Multiplying Eqs.~\eqref{eq:RWA1} and \eqref{eq:RWA2} yields the unified secular equation
\begin{equation}
(\omega-\omega_k^{(\sigma)})(\omega-\omega_P)-\sigma g^2=0,
\label{eq:secular}
\end{equation}
where the magnetoelastic coupling strength is $g^2=\gamma b^2\mu_0M_sk^2/(4\rho\omega_P)$. The sign $\sigma$ determines the hybridization topology. For $\sigma=+1$, Eq.~\eqref{eq:secular} yields the standard avoided crossing. For the SOT-stabilized antimagnon ($\sigma=-1$), the coupling term enters with a minus sign, giving \textit{level attraction} [Fig.~\ref{fig:fig1}(c)],
\begin{equation}
\omega_\pm=\omega_0\pm\sqrt{\frac{\Delta^2}{4}-g^2},
\label{eq:attraction}
\end{equation}
with $\omega_0=(\omega_P+\omega_k^{(a)})/2$ and $\Delta=\omega_P-\omega_k^{(a)}$. In the undamped/real-frequency approximation, the coalescence occurs at $|\Delta|=2g$~\cite{tserkovnyak_exceptional_2020,liu_observation_2019}. For $|\Delta|<2g$ the two branches form a complex-conjugate pair, while for $|\Delta|>2g$ they remain real. Strictly, when \(\Delta\) becomes complex due to finite damping, this coalescence condition generalizes to $\Delta^2=4g^2$. At the ideal crossing point $\Delta=0$, Eq.~\eqref{eq:attraction} yields $\omega_\pm=\omega_0\pm ig$, explicitly showing the complex-conjugate nature.

The physical origin of this sign reversal lies in the \textit{handedness flip} [Fig.~\ref{fig:fig1}(a)]. The positive-frequency resonance is carried by $m_+\equiv\delta m_x+i\delta m_y$ for both $\sigma=\pm1$. For $\sigma=+1$, $m_+$ is right-handed about $\mathbf m_0=+\hat{\mathbf{z}}$ and is the ordinary magnon. For $\sigma=-1$, the same $m_+$ is left-handed about $\mathbf m_0=-\hat{\mathbf{z}}$ and is the positive-frequency antimagnon. In the scalar RWA equations, this handedness flip manifests as the background magnetization $m_{0z}=\sigma$ entering the magnetoelastic coupling vertex, i.e., an intrinsic anti-phase channel, placing the factor $-\sigma g^2$ in the secular equation. When a positive-energy phonon hybridizes with the antimagnon, the dc spin current supplies energy to the coupled system. The modal amplitudes grow exponentially while both uncoupled modes decay, indicating a collective instability~\cite{wang_single-mode_2025}. We predict that this chirality argument directly leads to the observable level attraction. A detailed derivation is given in the SM~\cite{sm}.

\textit{Time-domain fingerprints and the SOT double-edged sword}---To expose the observable consequences of the momentum-space level attraction, we derive the time-domain envelope equations for a narrow wavepacket centered at the crossing wavevector $k_c$ (see SM~\cite{sm}). Writing the fields in envelope form around $k_c$, and expanding the bare frequencies to first order in $(k-k_c)$, the RWA equations become
\begin{align}
(\partial_t+v_k\partial_x+\Gamma_k)\delta m_{x,0}-g_k u_{z,0}&=0,
\label{eq:env1}\\
(\partial_t+v_p\partial_x+\Gamma_p)u_{z,0}+\sigma g_p\delta m_{x,0}&=0,
\label{eq:env2}
\end{align}
where $v_k=\sigma\,2\gamma Ak_c$, $v_p=c_t$, $g_k=\gamma bk_c/2$, $g_p=b_2k_c/(2\rho\omega_P)$, $g_kg_p=g^2$, $\Gamma_k=\sigma[\alpha\gamma(H+\sigma Ak_c^2)-J_{\mathrm{s}}]$ is the net magnetic damping, and $\Gamma_p$ is the phonon loss. 

For a uniform mode near the crossing point, the spatial gradients vanish. Seeking exponentially evolving solutions \(\propto e^{\lambda t}\) for the coupled envelope equations yields the complex growth rates (see SM~\cite{sm} for detailed derivation):
\begin{equation}
\lambda_\pm = -\frac{\Gamma_k + \Gamma_p}{2} \pm \sqrt{\left(\frac{\Gamma_k - \Gamma_p}{2}\right)^2 - \sigma g^2}.
\label{eq:growth}
\end{equation}
Physically, the real part \(\mathrm{Re}\,\lambda_\pm\) determines the temporal evolution of the mode amplitudes. The first term represents the average dissipation, while the square root captures the effect of coherent coupling. For the ordinary magnon (\(\sigma=+1\)), the coupling term \(-g^2\) maintains \(\mathrm{Re}\,\lambda_\pm < 0\), corresponding to stable decay (with negative \(\Gamma_k\) outside the antiparallel channel considered here). For the antimagnon (\(\sigma=-1\)), however, the intrinsic anti-phase channel reverses the sign to \(+g^2\). This level attraction allows the positive branch \(\lambda_+\) to overcome the mean dissipation and turn positive (\(\mathrm{Re}\,\lambda_+ > 0\)) when the coupling exceeds the net losses (\(g^2 > \Gamma_k\Gamma_p\)), signaling a collective instability that drives the exponential growth of the hybrid mode. Marginal stability corresponds to \(g^2 = \Gamma_k\Gamma_p\)~\cite{an_optimizing_2023}. 

Beyond this temporal growth dynamics (eigenvalues), the nature of the coupling is also imprinted on the eigenvector. The relative phase $\Delta\phi=\arg(u_z)-\arg(\delta m_x)$ distinguishes the two regimes [Fig.~\ref{fig:fig2}(b)--(d)]. From Eq.~\eqref{eq:RWA1}, \(u_z/\delta m_x=2i(\omega-\omega_k^{(\sigma)})/(\gamma bk)\). For level repulsion ($\sigma=+1$), the real eigenfrequencies ($\omega=\omega_0\pm g$) yield a purely imaginary eigenvector ratio, resulting in a relative phase of $\Delta\phi=\pm\pi/2$ (quadrature). Conversely, for level attraction ($\sigma=-1$), the complex-conjugate eigenfrequencies ($\omega=\omega_0\pm ig$) produce a purely real ratio. This locks the relative phase to \(0\) (in-phase) or \(\pi\) (anti-phase), reflecting a direct proportional response. Consequently, the magnetic and elastic components grow or decay synchronously as a single hybrid entity rather than exchanging energy periodically. This synchrony is confirmed by the in-phase oscillations and ballistic acoustic wave fronts in Figs.~\ref{fig:fig2}(b)--(d). Near the crossing point, this \(k\)-dependent \(0/\pi\) phase structure, dictated by broken time-reversal symmetry and the sign of \(k\), provides a clear experimental signature for time-resolved pump-probe measurements. Such a sharp \(0/\pi\) phase structure serves as a definitive fingerprint to distinguish our predicted intrinsic level attraction from the conventional \(\pi/2\) quadrature of level repulsion. The full phase analysis is given in the SM~\cite{sm}.

Here, the SOT rate $J_{\mathrm{s}}$ plays a counterintuitive role [Fig.~\ref{fig:fig2}(a)]. While $J_{\mathrm{s}}>\alpha\gamma H$ is strictly required to stabilize the antiparallel equilibrium and bring the antimagnon branch into existence (the lower threshold), the net magnetic damping $\Gamma_k=J_{\mathrm{s}}-\alpha\gamma(H-Ak^2)$ increases linearly with $J_{\mathrm{s}}$. Differentiating Eq.~\eqref{eq:growth} with respect to $\Gamma_k$, one finds
\begin{equation}
\frac{\partial\,\mathrm{Re}\,\lambda_+}{\partial\Gamma_k}
=-\frac12+\frac12\frac{(\Gamma_k-\Gamma_p)/2}{\sqrt{((\Gamma_k-\Gamma_p)/2)^2+g^2}}<0,
\label{eq:derivative}
\end{equation}
since the second term is strictly less than \(1/2\) for any $g>0$. Because $\partial\Gamma_k/\partial J_{\mathrm{s}}=1>0$, increasing $J_{\mathrm{s}}$ monotonically decreases the net growth rate. The SOT acts as a double-edged sword: it is the \textit{creator} of the antimagnon, but simultaneously the \textit{suppressor} of its self-sustained dynamics. This dual role fundamentally reflects the SOT-controlled balance between coherent gain and magnetic damping. Pushing $J_{\mathrm{s}}$ beyond the upper critical threshold
\begin{equation}
J_{\mathrm{s}}^{\mathrm{crit}}=\alpha\gamma(H-Ak^2)+\frac{g^2}{\Gamma_p},
\label{eq:Jscrit}
\end{equation}
drives the system into a regime where both hybrid modes decay, completely quenching the collective instability. Although this analytical threshold relies on the RWA (which breaks down at large \(J_s\), e.g., \(10^{11}~{\rm s^{-1}}\) in Fig. 2), our full time-domain simulations confirm the strong suppression. The critical phonon damping is \(\Gamma_p^{\rm crit} = g^2/[J_{\rm s} - \alpha\gamma(H - Ak^2)]\), and the growth window is \(\alpha\gamma H < J_{\rm s} < J_{\rm s}^{\rm crit}\). The lower bound corresponds to the linear stability limit of the uncoupled magnetic sector. A finite growth window requires \(g^2 > \alpha\gamma Ak^2\Gamma_p\), which is satisfied near the crossing for our parameters. The full derivation is provided in the SM~\cite{sm}.

\begin{figure}[t]
    \centering
    \includegraphics[width=\columnwidth]{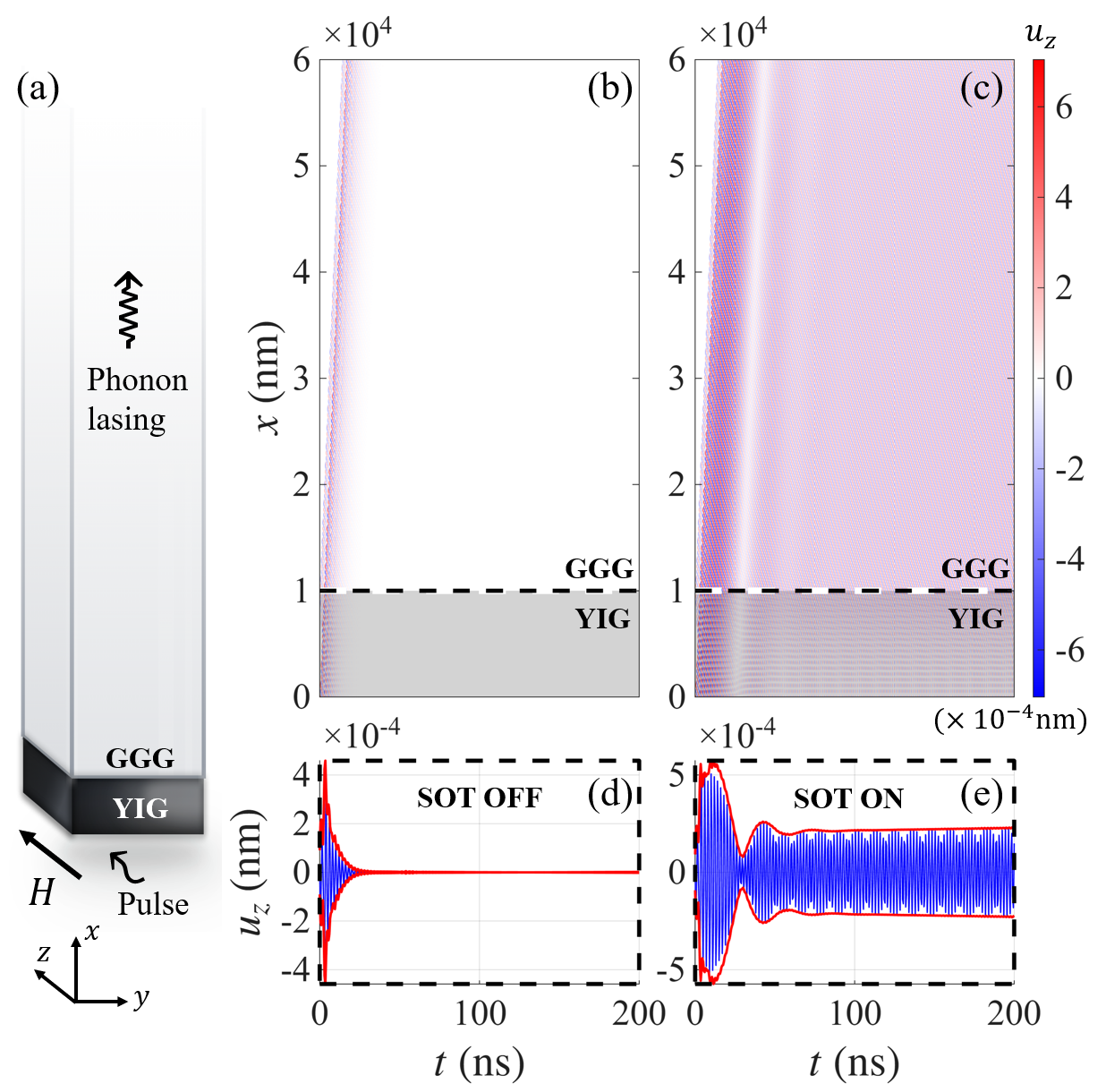}
    \caption{Proposed SOT-driven single-frequency phonon laser in a YIG/GGG heterostructure. (a) Schematic of the device. A YIG thin film of thickness \(d\) is in contact with a semi-infinite GGG substrate. A dc spin current exerts an SOT that stabilizes the antiparallel equilibrium, and a global field pulse along \(x\) initiates the magnetization dynamics. (b)–(e) Spatiotemporal evolution (b,c) and fixed-position time traces (d,e) of the transverse displacement \(u_z\). With the SOT off [(b) and (d)], the pulse-injected energy dissipates through phonon pumping into the substrate and intrinsic Gilbert damping, resulting in ballistic propagation and eventual decay. With the SOT on [(c) and (e)], the continuous energy injection from the SOT overcomes these losses, amplifying the acoustic field to establish a quasi-steady state that provides the gain mechanism for phonon lasing.}
    \label{fig:fig3}
\end{figure}

\textit{SOT-driven single-frequency phonon laser}---The momentum-space level attraction and the associated collective instability naturally enable a continuous-wave, single-frequency phonon laser~\cite{zhang_dissipative_2022,wang_single-mode_2025,bassant_phonon_2025}. We propose a device based on a YIG/GGG heterostructure [Fig.~\ref{fig:fig3}(a)]. The YIG film is in contact with a semi-infinite GGG substrate. We adopt GGG parameters $\rho_{\rm GGG}\simeq7.08\times10^{3}$~kg/m$^3$ and $c_{t,\rm GGG}\simeq3.5\times10^{3}$~m/s~\cite{clark_elastic_1961}. The GGG acts as both an acoustic waveguide and a phonon sink because its acoustic impedance closely matches that of YIG. At the YIG/GGG interface, the displacement and total shear traction are continuous: $u_z = u_z^{\rm sub}$ and $\mu\partial_x u_z + \sigma b_2 m_x = \mu_{\rm GGG}\partial_x u_z^{\rm sub}$ \cite{sato_dynamic_2021}. Crucially, this architecture requires no external high-frequency microwave pumping, in contrast to parametric magnon amplification~\cite{frostad_spin-transfer-assisted_2022,breitbach_stimulated_2023}. The dc current directly provides the energy for this single-frequency acoustic emission, and the highly wavevector-selective gain window suppresses other competing modes.

Because the crossing condition depends on the external static field \(H\), the output phonon frequency is continuously tunable by \(H\). The highly wavevector-selective gain window amplifies the initiated magnetization dynamics into a quasi-monochromatic, directional acoustic beam. The lasing is initiated by a global field pulse along \(x\) to rapidly excite the magnetization dynamics. Although the pulse is spatially uniform, the magnetoelastic boundary conditions break the translational symmetry, inevitably generating finite-\(k\) strain gradients that seed the propagating antimagnon-polaron modes. Consequently, the time trace in Fig.~\ref{fig:fig3}(e) also contains a background oscillation corresponding to the directly excited \(k=0\) Kittel mode. Our time-domain simulations of the YIG/GGG heterostructure confirm this physical picture (Fig.~\ref{fig:fig3}). With the SOT off, the injected acoustic energy dissipates through phonon pumping \cite{streib_damping_2018} into the substrate and intrinsic damping. With the SOT on, the continuous energy injection from the SOT overcomes these losses, amplifying the acoustic field to establish a steady envelope.

The lasing threshold is determined by the balance between the coupling gain and the total losses. The radiation of phonons into the GGG substrate provides an additional, and in practice dominant, loss channel compared to the intrinsic YIG attenuation \(\Gamma_p^{(0)}\). The effective phonon loss is $\Gamma_p^{\rm eff}=\Gamma_p^{(0)}+T\,c_{t,\rm YIG}/(2d)$, where \(T\) is the interfacial power transmissivity. Replacing \(\Gamma_p\to\Gamma_p^{\rm eff}\) in Eq.~\eqref{eq:Jscrit}, the radiation-corrected critical SOT rate becomes
\begin{equation}
J_{\mathrm{s}}^{\rm crit}
=\alpha\,\omega_{\rm out}
+\frac{g^{2}(k^*)}{\Gamma_p^{(0)}+T\,c_{t,\rm YIG}/(2d)}.
\label{eq:Jscrit_corrected}
\end{equation}
Taking an effective damping $\alpha=10^{-2}$, the theoretical lasing window is estimated as $2.2\times10^8\lesssim J_{\mathrm{s}}\lesssim3.8\times10^8$~s$^{-1}$ (see SM \cite{sm}). Our time-domain simulations yield a threshold of $2.89\times10^8$~s$^{-1}$, which is in good agreement with the analytic estimate. Experimentally, the same techniques used in the recent observation of antimagnons~\cite{wang_direct_2026}, such as Brillouin light scattering, can be directly applied here to detect the emitted acoustic signal and verify the predicted \(0/\pi\) phase relationship. 

\textit{Conclusion and discussion}---In summary, we have demonstrated that the antimagnon-polaron in a continuous magnetoelastic medium provides an intrinsic platform for coherent momentum-space level attraction, culminating in a continuously tunable phonon laser. Unlike discrete magnetic multilayers where such attraction relies on dipolar interactions~\cite{liu_tunable_2025}, our mechanism arises purely from the coherent magnetoelastic coupling mediated by the opposite-handedness of the antimagnon. We note that while this level attraction is intrinsic to the antimagnon-polaron, stabilizing the antimagnon itself requires an external dc SOT that effectively acts as a gain environment, which is conceptually consistent with the established understanding of environment-mediated dissipative coupling. It is worth noting that the negative-energy magnon-polaron discussed in Ref.~\cite{bassant_phonon_2025} corresponds, in the language of the present work, to the antimagnon-polaron predicted here. This analysis provides a highly integrated building block for future phononic circuits and microwave-to-acoustic transducers. The rapid experimental progress in observing antimagnons~\cite{wang_direct_2026} paves the way for verifying this level attraction via wave-vector-resolved Brillouin light scattering. Looking forward, in the supercritical regime, nonlinear saturation mechanisms such as four-magnon scattering and anharmonic phonon damping will ultimately dictate the steady-state lasing spectrum. A comprehensive nonlinear analysis of these effects in continuous magnetoelastic metamaterials~\cite{harms_antimagnonics_2024,bassant_phonon_2025,filatov_magnon-cherenkov_2026} will be addressed in future work.

\textit{Acknowledgments}--- This work was supported by Fundamental and Interdisciplinary Disciplines Breakthrough Plan of the Ministry of Education of China (Grant No. JYB2025XDXM120), National Key Research Program of China (Grant No. 2022YFA1403300), the Innovation Program for Quantum Science and Technology (Grant No. 2024ZD0300103) and National Natural Science Foundation of China (Grant No. 12574112 and 12204107). W. Yu thanks Z. Xia, A. L. Bassant and R. A. Duine for valuable discussions.

\bibliographystyle{apsrev4-2}
%

\clearpage
\onecolumngrid

\renewcommand{\thesection}{S\arabic{section}}
\section*{Supplemental Material for ``Prediction of antimagnon-polarons with intrinsic momentum-space level attraction''}
\label{app:sm}

\section{Full model and linearization to the unified secular equation}
\label{app:model}

The unit magnetization $\mathbf m=\mathbf M/M_s$ ($|\mathbf m|=1$) obeys the LLG equation augmented by the SOT,
\begin{equation}
\frac{\partial\mathbf m}{\partial t}=-\gamma\,\mathbf m\times\mathbf H_{\rm eff}
+\alpha\,\mathbf m\times\frac{\partial\mathbf m}{\partial t}
+J_{\mathrm{s}}\,\mathbf m\times(\mathbf p\times\mathbf m),
\label{eq:A-LLG}
\end{equation}
with $\mathbf p=-\hat{\mathbf{z}}$. The effective field (in A/m) contains exchange, the static external field, and the magnetoelastic contribution,
\begin{equation}
\mathbf H_{\rm eff}=A\nabla^2\mathbf m+\mathbf H_0+\mathbf H^{\rm mec},
\label{eq:A-heff}
\end{equation}
with $A$ the exchange coefficient and $\mathbf H_0=H\hat{\mathbf{z}}$ ($H>0$). Crystalline anisotropy and dipolar interaction are disregarded. The magnetoelastic field follows from the cubic energy density
\begin{equation}
U_{\rm mec}=b_1\sum_i m_i^2\varepsilon_{ii}+b_2\sum_{i\neq j}m_im_j\varepsilon_{ij},
\qquad
\varepsilon_{ij}=\tfrac12(\partial_iu_j+\partial_ju_i),
\label{eq:A-Ume}
\end{equation}
via $\mathbf H^{\rm mec}=-(\mu_0M_s)^{-1}\partial U_{\rm mec}/\partial\mathbf m$. The displacement obeys the elastic wave equation with the magnetoelastic body force,
\begin{equation}
\rho\frac{\partial^2\mathbf u}{\partial t^2}+\beta\frac{\partial\mathbf u}{\partial t}
=\mu\nabla^2\mathbf u+(\lambda+\mu)\nabla(\nabla\cdot\mathbf u)+\mathbf f^{\rm mec},
\label{eq:A-elastic}
\end{equation}
with $\mathbf f^{\rm mec}=\nabla\cdot\bar\Sigma$, $\Sigma_{ij}=\partial U_{\rm mec}/\partial(\partial_ju_i)$.

We consider the one-dimensional configuration with propagation along $x$ and equilibrium magnetization $\mathbf m_0=\sigma\hat{\mathbf z}$ ($\sigma=+1$ parallel, $\sigma=-1$ antiparallel). Setting $\mathbf H^{\rm mec}=0$ and writing $\mathbf m=\sigma\hat{\mathbf z}+\delta\mathbf m\,e^{i\omega t}$ with $\delta\mathbf m=(\delta m_x,\delta m_y,0)$, the linearization of Eq.~\eqref{eq:A-LLG} for a plane wave of wavevector $k$ gives, with $\omega_H\equiv\gamma H$ and $\omega_{\rm ex}(k)\equiv\gamma Ak^2$,
\begin{align}
i\omega\,\delta m_x&=-\big[\omega_H+\sigma\omega_{\rm ex}\big]\delta m_y-i\omega\alpha\sigma\,\delta m_y+\sigma J_{\mathrm{s}}\,\delta m_x,
\notag\\
i\omega\,\delta m_y&=+\big[\omega_H+\sigma\omega_{\rm ex}\big]\delta m_x+i\omega\alpha\sigma\,\delta m_x+\sigma J_{\mathrm{s}}\,\delta m_y .
\label{eq:A-linLLG}
\end{align}
Defining $\Omega_\sigma(k)\equiv\omega_H+\sigma\omega_{\rm ex}(k)$, the vanishing of the $2\times2$ determinant yields the exact secular equation
\begin{equation}
\big(i\omega-\sigma J_{\mathrm{s}}\big)^2+\big(\Omega_\sigma+i\omega\alpha\sigma\big)^2=0 .
\label{eq:A-sec_exact}
\end{equation}
Expanding to first order in the small parameters $\alpha$ and $J_{\mathrm{s}}/\Omega_\sigma$ gives the two branches
\begin{equation}
\omega_\pm^{(\sigma)}\simeq\pm\Omega_\sigma(k)+i\sigma\big[\alpha\Omega_\sigma(k)-J_{\mathrm{s}}\big].
\label{eq:A-branches}
\end{equation}
For $\sigma=+1$, the positive-frequency branch is the ordinary magnon,
\begin{equation}
\omega_k^{(m)}\simeq+\gamma(H+Ak^2)+i\big[\alpha\gamma(H+Ak^2)-J_{\mathrm{s}}\big],
\label{eq:A-bare_m}
\end{equation}
while for $\sigma=-1$, the positive-frequency branch is the antimagnon,
\begin{equation}
\omega_k^{(a)}\simeq+\gamma(H-Ak^2)+i\big[J_{\mathrm{s}}-\alpha\gamma(H-Ak^2)\big].
\label{eq:A-bare_a}
\end{equation}
Within the positive-frequency antimagnon branch, \(\Omega_{-1}(k)=\gamma(H-Ak^2)>0\) (i.e., \(|k|<\sqrt{H/A}\)); there the uncoupled antiparallel state is globally stable for \(J_{\mathrm{s}}>J_{\mathrm{s}}^{\rm st}\equiv\alpha\gamma H\), with the most stringent condition occurring at \(k=0\). At the crossing wavevector \(k^*\), the weaker local condition is \(J_{\mathrm{s}}>\alpha\gamma(H-Ak^{*2})\). For \(k>0\), the acoustic line \(\omega_P=c_tk\) crosses the antimagnon branch at the unique positive wavevector \(k^*\) solving \(c_tk^*=\gamma(H-Ak^{*2})\).

Linearizing the full coupled system around $\mathbf m_0=\sigma\hat{\mathbf{z}}$ and using the phasor convention, the nonzero strains are $\varepsilon_{xx}=\partial_xu_x$, $\varepsilon_{xy}=\tfrac12\partial_xu_y$, and $\varepsilon_{xz}=\tfrac12\partial_xu_z$. With all fields depending on $x$ only, the magnetoelastic energy density reduces to $U_{\rm mec}=b_1m_x^2\partial_xu_x+b_2m_xm_y\partial_xu_y+b_2m_zm_x\partial_xu_z$. Keeping only terms linear in the dynamical variables, the $b_1$ term and the $u_y$-related term yield third-order energies. Consequently, the effective fields and forces derived from them are of second order and are neglected in the linearized equations. The only surviving linear coupling is the \(b_2m_zm_x\partial_xu_z\) term, with \(m_z=\sigma\sqrt{1-\delta m_x^2-\delta m_y^2}\approx\sigma\) to leading order in the transverse fluctuations. The linearized magnetoelastic field and body force are
\begin{equation}
\mathbf H^{\rm mec}=-\sigma\frac{b_2}{\mu_0M_s}(\partial_xu_z)\hat{\mathbf{x}},
\qquad
\mathbf f^{\rm mec}=\sigma b_2(\partial_x\delta m_x)\hat{\mathbf{z}} .
\label{eq:A-lin_mec}
\end{equation}
Adding the magnetoelastic torque and using $\partial_x\to -ik$, the spin sector reads
\begin{align}
i\omega\,\delta m_x&=-\Omega_\sigma\delta m_y-i\omega\alpha\sigma\,\delta m_y+\sigma J_{\mathrm{s}}\,\delta m_x,
\notag\\
i\omega\,\delta m_y&=+\Omega_\sigma\delta m_x+i\omega\alpha\sigma\,\delta m_x+\sigma J_{\mathrm{s}}\,\delta m_y-i\gamma bk\,u_z ,
\label{eq:A-spincomp}
\end{align}
while the elastic sector reads
\begin{equation}
(\omega_P^2-\omega^2)u_z=-i\sigma\frac{b_2}{\rho}k\,\delta m_x ,
\label{eq:A-elastcomp}
\end{equation}
with $\omega_P=c_tk$ and $c_t=\sqrt{\mu/\rho}$. Eliminating $\delta m_y$ and applying the RWA near $\omega\approx\Omega_\sigma$, we obtain
\begin{equation}
(\omega-\omega_k^{(\sigma)})\delta m_x=-\frac{i\gamma bk}{2}u_z,
\qquad
(\omega-\omega_P)u_z=\frac{i\sigma b_2k}{2\rho\omega_P}\delta m_x .
\label{eq:A-RWA}
\end{equation}
Multiplying the two equations yields the unified secular equation
\begin{equation}
(\omega-\omega_k^{(\sigma)})(\omega-\omega_P)-\sigma g^2=0,
\qquad
g^2\equiv\frac{\gamma\,b^2\,\mu_0M_s\,k^2}{4\rho\,\omega_P}.
\label{eq:A-sec_unified}
\end{equation}
For $\sigma=+1$, the eigenvalues are real at the crossing, giving level repulsion. For $\sigma=-1$, the off-diagonal signs are opposite, giving level attraction:
\begin{equation}
\omega_\pm=\omega_0\pm\sqrt{\frac{\Delta^2}{4}-g^2},
\qquad
\omega_0=\frac{\omega_P+\omega_k^{(a)}}{2},\quad \Delta=\omega_P-\omega_k^{(a)} .
\label{eq:A-attraction}
\end{equation}
In the undamped/real-frequency approximation, the coalescence occurs at $|\Delta|=2g$. For $|\Delta|<2g$ the branches form a complex-conjugate pair, while for $|\Delta|>2g$ they remain real. Strictly, when $\Delta$ is complex, the square root vanishes when \(\Delta^2=4g^2\).

\section{Handedness flip and the origin of the sign reversal}
\label{app:chirality}

To expose the physics of the factor $\sigma$ in Eq.~\eqref{eq:A-sec_unified}, we introduce the two fixed circular amplitudes $m_\pm\equiv\delta m_x\pm i\delta m_y$ (lab-frame right-/left-handed). Under the $e^{i\omega t}$ convention, the positive-frequency resonance is carried by $m_+$ for \emph{both} $\sigma=\pm1$: for $\sigma=+1$, $m_+$ is right-handed about $\mathbf m_0=+\hat{\mathbf{z}}$ and is the ordinary magnon. For $\sigma=-1$, the same $m_+$ is left-handed about $\mathbf m_0=-\hat{\mathbf{z}}$ and is the positive-frequency antimagnon~\cite{zhou_left-handed_2019}, while the right-handed-about-$\mathbf m_0$ amplitude $m_-$ is pushed to the negative-frequency branch. Thus the opposite-handedness of the antimagnon and the handedness flip of the resonant positive-frequency amplitude are two sides of the same fact. In the scalar RWA equations, this handedness flip manifests as the background magnetization $m_{0z}=\sigma$ entering the magnetoelastic coupling vertex, and it is this $\sigma$ that places the factor $-\sigma g^2$ in the secular equation. In the language of the Bogoliubov-de Gennes formalism~\cite{harms_antimagnonics_2024}, this classical handedness flip corresponds to the antimagnon carrying a \emph{negative norm}. Our coupling-phase formulation is consistent with that picture but does not rely on it. When a positive-energy phonon hybridizes with the antimagnon, the dc spin current supplies energy to the coupled system. The modal amplitudes grow exponentially while both uncoupled modes decay, the fundamental origin of the \emph{collective instability}.

\section{Time-domain envelope and phase structure}
\label{app:envelope}

We derive the time-domain envelope equations from the frequency-domain RWA equations. We focus on a narrow wavepacket centered at the carrier wavevector $k_c$ and carrier frequency $\omega_c\equiv\mathrm{Re}[\omega_k^{(\sigma)}(k_c)]=\gamma(H+\sigma Ak_c^2)$. The full solutions are written as
\begin{equation}
\delta m_x(x,t)=\delta m_{x,0}(x,t)e^{i(\omega_ct-k_cx)},\qquad
u_z(x,t)=u_{z,0}(x,t)e^{i(\omega_ct-k_cx)},
\label{eq:A-ansatz}
\end{equation}
where the envelopes vary slowly in space and time. Expanding the bare frequencies to first order in $(k-k_c)$,
\begin{align}
\omega_k^{(\sigma)}(k)&\approx\omega_k^{(\sigma)}(k_c)+v_k(k-k_c),
\qquad v_k\equiv\sigma\,2\gamma Ak_c,
\label{eq:A-expand_m}\\
\omega_P(k)&\approx\omega_P(k_c)+v_p(k-k_c),
\qquad v_p\equiv c_t,
\label{eq:A-expand_p}
\end{align}
Under the slowly varying envelope approximation (SVEA), we replace \(\omega-\omega_c\to -i\partial_t\) and \(k-k_c\to i\partial_x\) acting on the envelopes \(\delta m_{x,0}\) and \(u_{z,0}\), and approximate \(k\approx k_c\) in the coupling coefficients. We obtain
\begin{align}
(\partial_t+v_k\partial_x+\Gamma_k)\,\delta m_{x,0}-g_k\,u_{z,0}&=0,
\label{eq:A-env_m}\\
(\partial_t+v_p\partial_x+\Gamma_p)\,u_{z,0}+\sigma g_p\,\delta m_{x,0}&=0,
\label{eq:A-env_p}
\end{align}
where the real coupling coefficients are
\begin{equation}
g_k\equiv\frac{\gamma bk_c}{2},\qquad
g_p\equiv\frac{b_2k_c}{2\rho\omega_P(k_c)},\qquad
g_kg_p=g^2,
\label{eq:A-couplings}
\end{equation}
and $\Gamma_k\equiv\sigma(\alpha\gamma(H+\sigma Ak_c^2)-J_{\mathrm{s}})$ is the net magnetic damping, $\Gamma_p$ the phonon damping. For a uniform mode ($q=0$), the growth rate is
\begin{equation}
\lambda_\pm=-\frac{\Gamma_k+\Gamma_p}{2}\pm\sqrt{\left(\frac{\Gamma_k-\Gamma_p}{2}\right)^2-\sigma g^2}.
\label{eq:A-lambda}
\end{equation}
For $\sigma=+1$ with positive dampings, $\mathrm{Re}\,\lambda_\pm<0$, so the mode is stable. For $\sigma=-1$, the square root becomes $\sqrt{((\Gamma_k-\Gamma_p)/2)^2+g^2}$, always real and positive, and the dominant eigenvalue is
\begin{equation}
\mathrm{Re}\,\lambda_+=-\frac{\Gamma_k+\Gamma_p}{2}+\sqrt{\left(\frac{\Gamma_k-\Gamma_p}{2}\right)^2+g^2}.
\label{eq:A-growth}
\end{equation}
Growth occurs when $g^2>\Gamma_k\Gamma_p$. Marginal stability at $g^2=\Gamma_k\Gamma_p$. Decay when $g^2<\Gamma_k\Gamma_p$.

The phase relationship between the magnon and phonon components follows from the eigenvector structure of the RWA equations. From Eq.~\eqref{eq:A-RWA},
\begin{equation}
u_z=\frac{2i(\omega-\omega_k^{(\sigma)})}{\gamma bk}\delta m_x .
\label{eq:A-evec_phase}
\end{equation}
We define the relative phase as $\Delta\phi\equiv\arg(u_z)-\arg(\delta m_x)$.

\textbf{Level repulsion ($\sigma=+1$).} At the crossing point $\omega_k^{(m)}=\omega_P=\omega_0$, the eigenfrequencies are real, $\omega-\omega_0=\pm g$ with $g>0$. The ratio is purely imaginary, and the relative phase is $\pm\pi/2$: the two components are in quadrature.

\textbf{Level attraction ($\sigma=-1$).} At the crossing point $\omega_k^{(a)}=\omega_P=\omega_0$, the eigenfrequencies are complex, $\omega-\omega_0=\pm ig$. The ratio is real, so the relative phase is either $0$ (in-phase) or $\pi$ (anti-phase). With the \(e^{i\omega t}\) convention, the growing root is \(\omega=\omega_0-ig\); it has \(\Delta\phi=0\) for \(k>0\), while the decaying root \(\omega=\omega_0+ig\) has \(\Delta\phi=\pi\). For \(k<0\), the ratio \(u_z/\delta m_x\) reverses sign, so the growing mode acquires \(\Delta\phi=\pi\) (anti-phase) while the decaying mode has \(\Delta\phi=0\). Thus level attraction is characterized by simultaneous growth (or decay) of both components with no oscillatory phase relation.

\textbf{Coalescence.} Strictly, the square root vanishes when $\Delta^2=4g^2$. In the undamped/real-frequency approximation this reduces to $|\Delta|=2g$, where the two eigenvalues coalesce. The $0/\pi$ phase structure discussed above applies at the \emph{crossing point} $\Delta=0$. At exact resonance the eigenvector ratio is real for any $\Gamma_k,\Gamma_p$. Detuning shifts the relative phase away from $0/\pi$. The phase structure is sharp near $\Delta=0$.

\section{Non-Hermitian dynamics and the SOT double-edged sword}
\label{app:Js}

The stability of the antimagnon-polaron is governed by the dominant eigenvalue $\lambda_+$. To determine the dependence on $J_{\mathrm{s}}$, we take the partial derivative of Eq.~\eqref{eq:A-growth} with respect to $\Gamma_k$ (which is linear in $J_{\mathrm{s}}$):
\begin{equation}
\frac{\partial\,\mathrm{Re}\,\lambda_+}{\partial\Gamma_k}
=-\frac12+\frac12\frac{(\Gamma_k-\Gamma_p)/2}{\sqrt{((\Gamma_k-\Gamma_p)/2)^2+g^2}} .
\label{eq:A-derivative}
\end{equation}
Letting $x=(\Gamma_k-\Gamma_p)/2$, the second term becomes $\frac12 x/\sqrt{x^2+g^2}$. Since \(g>0\), \(\displaystyle \frac{x}{\sqrt{x^2+g^2}}\le \frac{|x|}{\sqrt{x^2+g^2}}<1\), which proves \(\partial\,\mathrm{Re}\,\lambda_+/\partial\Gamma_k<0\) universally. Since $\partial\Gamma_k/\partial J_{\mathrm{s}}=1>0$, increasing $J_{\mathrm{s}}$ monotonically decreases the net growth rate. When $J_{\mathrm{s}}$ exceeds the upper critical threshold
\begin{equation}
J_{\mathrm{s}}^{\rm crit}(k)=\alpha\gamma(H-Ak^2)+\frac{g^2(k)}{\Gamma_p},
\label{eq:A-Jscrit}
\end{equation}
both hybrid modes decay, and the collective instability is destroyed, so that self-sustained growth is completely quenched. Equivalently, the critical phonon damping is
\begin{equation}
\Gamma_p^{\rm crit}=\frac{g^2}{J_{\mathrm{s}}-\alpha\gamma(H-Ak^2)},
\label{eq:A-Gpcrit}
\end{equation}
and the growth window is
\begin{equation}
\alpha\gamma H<J_{\mathrm{s}}<\alpha\gamma(H-Ak^2)+\frac{g^2}{\Gamma_p}.
\label{eq:A-Jswindow}
\end{equation}
We note that when \(J_s\) becomes comparable to or larger than the carrier frequency (e.g., \(J_s=10^{11}~{\rm s^{-1}}\)), the RWA ceases to be valid. However, full numerical simulations of Eqs.~\eqref{eq:A-LLG}-\eqref{eq:A-elastic} show that the suppressed regime persists qualitatively in this strongly dissipative limit.

\section{Phonon radiation loss and the corrected lasing threshold in the YIG/GGG heterostructure}
\label{app:phonon_pump}

The lasing threshold Eq.~\eqref{eq:Jscrit} contains the phonon damping $\Gamma_p$, which in a bulk one-dimensional model is the intrinsic YIG phonon attenuation $\Gamma_p^{(0)}$. In the YIG/GGG heterostructure, however, the YIG film is in contact with a semi-infinite GGG substrate that acts as a phonon sink, and the radiation of phonons into the substrate provides an additional, and in practice dominant, loss channel. We derive this radiation loss and the corrected threshold. The total effective phonon damping is \(\Gamma_p^{\rm eff}=\Gamma_p^{(0)}+T\,c_{t,\rm YIG}/(2d)\), and replacing \(\Gamma_p\to\Gamma_p^{\rm eff}\) in Eq.~\eqref{eq:A-Jscrit} yields Eq.~\eqref{eq:Jscrit_corrected} of the main text.

\textbf{Geometry and acoustic impedances.} Let the YIG film occupy \(0<x<d\), with the free surface at $x=0$ and the YIG/GGG interface at \(x=d\). The GGG substrate occupies \(x>d\). Denote the YIG density and transverse sound velocity by $\rho_{\rm YIG},c_{t,\rm YIG}$ and those of GGG by $\rho_{\rm GGG},c_{t,\rm GGG}$. The acoustic impedances are
\begin{equation}
Z_{\rm YIG}=\rho_{\rm YIG}\,c_{t,\rm YIG},
\qquad
Z_{\rm GGG}=\rho_{\rm GGG}\,c_{t,\rm GGG}.
\end{equation}

\textbf{Interfacial reflection and transmission.} In the purely elastic limit, i.e., neglecting the magnetoelastic contribution \(\sigma b_2 m_x\) to the shear traction at the interface, matching displacement and shear stress at \(x=d\) gives the amplitude reflection and transmission coefficients for a wave incident from the YIG side,
\begin{equation}
r=\frac{Z_{\rm YIG}-Z_{\rm GGG}}{Z_{\rm YIG}+Z_{\rm GGG}},
\qquad
t=\frac{2Z_{\rm YIG}}{Z_{\rm YIG}+Z_{\rm GGG}}.
\end{equation}
The power transmissivity across the interface is
\begin{equation}
T=1-|r|^2=\frac{4Z_{\rm YIG}Z_{\rm GGG}}{(Z_{\rm YIG}+Z_{\rm GGG})^2}.
\label{eq:A-T}
\end{equation}
With the YIG parameters $Z_{\rm YIG}=5170\times3783\approx1.96\times10^{7}$~kg\,m$^{-2}$s$^{-1}$ and GGG parameters $Z_{\rm GGG}=7080\times3500\approx2.48\times10^{7}$~kg\,m$^{-2}$s$^{-1}$, we obtain $T\approx0.986$, i.e., the interface is nearly acoustically transparent and $98.6\%$ of the incident phonon energy escapes into the substrate.

\textbf{Radiation loss rate.} The ratio of the acoustic power radiated into GGG to the elastic energy stored in the YIG film defines an \emph{energy} decay rate. For a monochromatic mode of frequency $\omega$, the time-averaged energy density in the film is $\tfrac12\rho_{\rm YIG}\omega^2|u|^{2}$, so the stored energy per unit area is $E_{\rm stored}=\tfrac12\rho_{\rm YIG}\omega^2|u|^{2}d$. The power radiated into the substrate per unit area is $P_{\rm out}=\tfrac12Z_{\rm GGG}\omega^2|u_t|^{2}$, with $u_t=t\,u$. Hence the energy decay rate is
\begin{equation}
\Gamma_E
=\frac{P_{\rm out}}{E_{\rm stored}}
=\frac{Z_{\rm GGG}\,|t|^{2}}{\rho_{\rm YIG}\,d}
=\frac{4Z_{\rm YIG}Z_{\rm GGG}}{(Z_{\rm YIG}+Z_{\rm GGG})^{2}}\cdot\frac{c_{t,\rm YIG}}{d}
=T\,\frac{c_{t,\rm YIG}}{d}.
\label{eq:A-Gamma_rad}
\end{equation}
Because the stored energy scales as the square of the field amplitude, an amplitude envelope decaying as $e^{-\Gamma_p t}$ carries an energy decaying as $e^{-2\Gamma_p t}$. The amplitude decay rate entering the envelope equations is therefore half the energy decay rate,
\begin{equation}
\Gamma_p^{\rm rad}=\frac{\Gamma_E}{2}=\frac{P_{\rm out}}{2E_{\rm stored}}=\frac{T}{2}\,\frac{c_{t,\rm YIG}}{d}.
\end{equation}
For a nearly transparent interface $T\approx1$, this reduces to the physical estimate $\Gamma_p^{\rm rad}\approx c_{t,\rm YIG}/(2d)$: the escape time is set by the phonon transit time across the film. For $d=10^{4}$~nm and $c_{t,\rm YIG}=3783$~m/s, we obtain
\begin{equation}
\Gamma_p^{\rm rad}=0.986\times\frac{3783}{2\times10^{-5}}\approx1.87\times10^{8}~\mathrm{s^{-1}},
\end{equation}
which is roughly \(20\) times larger than the intrinsic YIG attenuation \(\Gamma_p^{(0)}=\beta/(2\rho)\approx9.5\times10^{6}~\mathrm{s^{-1}}\).

\textbf{Corrected lasing threshold.} With \(\Gamma_p^{\rm eff}=\Gamma_p^{(0)}+T\,c_{t,\rm YIG}/(2d)\), and using \(\gamma(H-Ak^{*2})=\omega_P(k^*)=c_{t,\rm YIG}k^*\equiv\omega_{\rm out}\) at the crossing wavevector \(k^*\), replacing \(\Gamma_p\to\Gamma_p^{\rm eff}\) in Eq.~\eqref{eq:A-Jscrit} yields the corrected critical SOT rate Eq.~\eqref{eq:Jscrit_corrected} of the main text.

\textbf{Numerical evaluation.} Using the YIG parameters listed in the main text, the antimagnon-phonon crossing wavevector is $k^*\approx5.66\times10^{6}$~m$^{-1}$, giving $\omega_{\rm out}=c_{t,\rm YIG}k^*\approx2.14\times10^{10}$~rad/s and $g^{2}(k^*)\approx3.17\times10^{16}$. Taking $\alpha=10^{-2}$, the intrinsic Gilbert contribution is $\alpha\omega_{\rm out}\approx2.14\times10^{8}$~s$^{-1}$. The effective phonon damping is $\Gamma_p^{\rm eff}\approx9.5\times10^{6}+1.87\times10^{8}\approx1.96\times10^{8}$~s$^{-1}$, so that $g^{2}/\Gamma_p^{\rm eff}\approx1.6\times10^{8}$~s$^{-1}$ and $J_{\mathrm{s}}^{\rm crit}\approx2.14\times10^{8}+1.6\times10^{8}\approx3.8\times10^{8}$~s$^{-1}$. This simple estimate is of the same order as the value $2.89\times10^{8}$~s$^{-1}$ obtained from time-domain simulations of the YIG/GGG heterostructure, which we adopt as the operative threshold in the device discussion of Fig.~\ref{fig:fig3}, whereas ignoring the radiation loss would give $3.55\times10^{9}$~s$^{-1}$, an order of magnitude too large. We conclude that phonon radiation into the GGG substrate dominates the phonon loss channel and is essential for a quantitative prediction of the lasing threshold.

\textbf{Validity of the storage approximation.} The radiation-loss estimate above treats the film field as a uniformly stored travelling wave and uses the plane-wave power transmissivity $T$. Because $T\approx0.986$, the interface is nearly transparent and the film acts as a strongly leaky, low-quality-factor resonator rather than a well-confined acoustic cavity. The uniform-storage assumption is therefore only a same-order estimate. A rigorous radiation loss would be extracted from the imaginary part of the complex frequency of the open-boundary elastic eigenproblem, which in the strong-leakage limit differs from $Tc_{t,\rm YIG}/(2d)$ by an $O(1)$ factor and accounts for the residual difference between the analytic $3.8\times10^{8}$~s$^{-1}$ and the simulated $2.89\times10^{8}$~s$^{-1}$. A full complex-frequency eigenmode analysis of the open-boundary heterostructure would refine this \(O(1)\) prefactor but is beyond the scope of this SM. We therefore adopt the numerically obtained threshold \(J_{\mathrm{s}}^{\rm crit}\approx2.89\times10^{8}~\mathrm{s^{-1}}\) for the device discussion in the main text.

\section{Numerical simulation details}
\label{app:num}

For the one-dimensional model of Figs.~\ref{fig:fig1} and \ref{fig:fig2}, the film is taken along $x$. The YIG/GGG heterostructure in Fig.~\ref{fig:fig3} is treated as a guided-mode problem with the GGG substrate acting as an additional phonon sink, as described in SM, Sec.~\ref{app:phonon_pump}.

This appendix summarizes the numerical implementation. All time-domain and frequency-domain micromagnetic simulations are performed using the Micromagnetics Module of COMSOL Multiphysics~\cite{comsol,Zhang2023,wang_effective_2026}. The time-domain simulations solve the full LLG equation [Eq.~\eqref{eq:LLG}] coupled to the elastic wave equation [Eq.~\eqref{eq:A-elastic}] with the magnetoelastic body force derived in Sec.~\ref{app:model}. The magnetic film is modeled as a 1D chain along the propagation direction $x \in [0, d]$, with a semi-infinite GGG substrate attached at \(x=d\). The finite length \(d\) along the propagation direction is incorporated analytically via the effective radiation loss derived in SM, Sec.~\ref{app:phonon_pump}.

\textbf{Time-domain magnetoelastic boundary conditions.} At the free surface $x=0$, the total shear traction vanishes, $\mu\partial_xu_z+\sigma b_2m_x=0$. At the film-substrate interface $x=d$, displacement and total shear traction are continuous, $u_z^{\rm film}=u_z^{\rm sub}$ and $\mu\partial_xu_z^{\rm film}+\sigma b_2m_x=\mu_{\rm sub}\partial_xu_z^{\rm sub}$. The substrate is terminated by a perfectly matched layer to absorb outgoing phonons. For the magnetization, free exchange boundary conditions $\partial_x\mathbf m=0$ are used. The SOT is applied uniformly. The Kittel mode is excited by a global external field pulse along \(x\), $\mathbf H_{\rm pulse}(t)=H_{\rm pulse}\hat{\mathbf{x}} f(t)$, with $f(t)$ a short Gaussian or sinc pulse whose bandwidth covers the Kittel frequency. After the pulse, the free evolution of $|\delta m_x(t)|$ and $|u_z(t)|$ at the film center is monitored. We note that although the external pulse is spatially uniform (\(k=0\)), the magnetoelastic boundary conditions at the free surface and the interface break translational symmetry, converting the uniform magnetization dynamics into finite-\(k\) strain gradients that seed the propagating antimagnon-polaron modes.

\textbf{Frequency-domain linearization.} The frequency-domain solver uses the phasor convention $\propto e^{i\omega t}$. Writing $\mathbf m=\mathbf m_0+\delta\mathbf m e^{i\omega t}$ and $\mathbf u=\delta\mathbf u e^{i\omega t}$ with $\mathbf m_0=\sigma\hat{\mathbf{z}}$, and keeping terms linear in the dynamical variables, the LLG equation becomes
\begin{equation}
i\omega\,\delta\mathbf m
=-\gamma\,\mathbf m_0\times\delta\mathbf h^{\rm eff}
-\gamma\,\delta\mathbf m\times\mathbf h_0^{\rm eff}
+i\omega\alpha\,\mathbf m_0\times\delta\mathbf m,
\end{equation}
where \(\delta\mathbf h^{\rm eff}=A\nabla^2\delta\mathbf m+\delta\mathbf h^{\rm mec}+\delta\mathbf h^{\rm SOT}\) and \(\mathbf h_0^{\rm eff}=A\nabla^2\mathbf m_0+\mathbf H_0\). Here \(\delta\mathbf h^{\rm SOT}\equiv (J_{\mathrm{s}}/\gamma)\hat{\mathbf z}\times\delta\mathbf m\) denotes the linear effective field representing the SOT torque for \(\mathbf p=-\hat{\mathbf z}\) and \(\mathbf m_0=\sigma\hat{\mathbf z}\). For the frequency-domain dispersion calculation we set \(\beta=0\), so that the linearized elastic equation reads
\begin{equation}
-\rho\omega^2\delta\mathbf u
=\mu\nabla^2\delta\mathbf u+(\lambda+\mu)\nabla(\nabla\cdot\delta\mathbf u)
+\delta\mathbf f^{\rm mec},
\end{equation}
with the magnetoelastic field and body force given in Eqs.~\eqref{eq:A-lin_mec} and the surrounding text. In the one-dimensional geometry, the only coupled channel for $\mathbf m_0=(0,0,-1)$ is $(\delta m_x,\delta u_z)$, and the equations reduce to Eqs.~\eqref{eq:A-spincomp}--\eqref{eq:A-elastcomp}. The same mechanical boundary conditions as in the time domain are applied. The frequency is swept at fixed $H$, and the coalescence of the hybridized branches is tracked as a function of $k$ and $H$.

\end{document}